# Frequency moments of the Coulomb dynamic structure factor and related integrals

**B J B Crowley**[1,2]

[1]Department of Physics, University of Oxford, Parks Road, Oxford OX1 3PU, UK

[2]Email: *basil.crowley@physics.ox.ac.uk*

Date: 25 December 2015

This report addresses the moments, $\mathcal{S}_n(\mathbf{q}) = \int_{-\infty}^{+\infty} \omega^n S(\mathbf{q},\omega) \, d\omega \quad n \in \mathbb{N}, \ n \geq -1$, of the quantum mechanical dynamic structure factor $S(\mathbf{q},\omega)$ for a one-component Coulomb plasma in thermodynamic equilibrium. The Fluctuation Dissipation Theorem relates these moments to integrals involving the imaginary part of the inverse longitudinal dielectric function, with the odd moments in particular being equivalent to the odd moments of the imaginary part of the inverse dielectric function. Application of the Generalized Plasmon Pole Approximation [1] to a weakly-coupled non-degenerate plasma, leads to general formulae expressed in terms of polynomial functions. Explicit forms of these functions are given for $n \leq 20$. These formulae are generalized to degenerate and partially degenerate plasmas, in small-**q** (long-wavelength) regimes.

This page is intentionally left blank

**CONTENTS**

This page is intentionally left blank

# 1 INTRODUCTION

## 1.1 Introduction

Consider a one-component *Coulomb* plasma with Hamiltonian $\hat{H}$ comprising particles of mass $m$ and charge $e$, in thermodynamic equilibrium at temperature $T$ and mean density $n$. Here, units are used throughout such that Planck's constant $\hbar$ and Boltzmann's constant $k$ both have the value of unity, resulting in temperature, energy and frequency being expressed in the same units - typically electron volts (eV).

The *dynamic structure factor* [2], [3],

$$S(\mathbf{q},\omega) = \frac{1}{2\pi n \mathfrak{V}} \int_{-\infty}^{+\infty} \langle \hat{\rho}_{-\mathbf{q}}(\tfrac{1}{2}t) \hat{\rho}_{\mathbf{q}}(-\tfrac{1}{2}t) \rangle e^{i\omega t} dt \qquad (1)$$

is a two-body *correlation function* equivalent to the spatial and temporal Fourier transform of the density-density correlation function, $\langle \hat{\rho}_{\mathbf{r'}+\mathbf{r}/2}(\tfrac{1}{2}t) \hat{\rho}_{\mathbf{r'}-\mathbf{r}/2}(-\tfrac{1}{2}t) \rangle$. It plays a central role in the treatment of scattering and equations of state of many body systems in local thermodynamic equilibrium. In equation (1), $\mathfrak{V}$ is the volume, $\hat{\rho}_{\mathbf{q}} = \int_{\mathfrak{V}} \hat{\rho}_{\mathbf{r}} e^{-i\mathbf{q}\cdot\mathbf{r}} d^3\mathbf{r}$ is the density operator, and $\langle \hat{O} \rangle = \text{Tr}(\hat{O} e^{-\hat{H}/T})$, denotes the thermal average or expectation value of an observable $\hat{O}$.

The function (1) satisfies a number of important general theorems [1], [4], [5], [6]. These include a detailed-balance relation of the *Kubo-Martin-Schwinger* (KMS) type,

$$S(\mathbf{q},-\omega) = e^{-\omega/T} S(\mathbf{q},\omega) \qquad (2)$$

and the *Fluctuation-Dissipation Theorem*,

$$S(\mathbf{q},\omega) = \frac{q^2}{\pi m \Omega_0^2} \frac{1}{1-e^{-\omega/T}} \text{Im}\left(\frac{-1}{\epsilon(\mathbf{q},\omega)}\right) \qquad (3)$$

where

$$\Omega_0 = \sqrt{\frac{ne^2}{\epsilon_0 m}} \qquad (4)$$

is the *plasma frequency*, and $\epsilon(\mathbf{q},\omega)$ is the longitudinal *dielectric function*.

The frequency moments $\mathfrak{S}_k(\mathbf{q}) = \int_{-\infty}^{+\infty} \omega^k S(\mathbf{q},\omega) \, d\omega$ of (1) yield the time derivatives of the density autocorrelation function according to

$$n\mathfrak{V}\mathfrak{S}_k(\mathbf{q}) = i^k \frac{\partial^k}{\partial t^k} \langle \hat{\rho}_{-\mathbf{q}}(\tfrac{1}{2}t) \hat{\rho}_{\mathbf{q}}(-\tfrac{1}{2}t) \rangle \Big|_{t=0} \quad (5)$$

The zeroth and first frequency moments are given respectively by the elastic sum rule,

$$\mathfrak{S}_0(\mathbf{q}) \equiv \int_{-\infty}^{\infty} S(\mathbf{q},\omega) \, d\omega = S(\mathbf{q}) \quad (6)$$

and the f-sum rule,

$$\mathfrak{S}_1(\mathbf{q}) \equiv \int_{-\infty}^{\infty} \omega S(\mathbf{q},\omega) \, d\omega = \frac{q^2}{2m} \quad (7)$$

In equation (6), $S(\mathbf{q})$ denotes the *static structure factor*,

$$S(\mathbf{q}) = \frac{1}{n\mathfrak{V}} \langle \hat{\rho}_{\mathbf{q}} \hat{\rho}_{-\mathbf{q}} \rangle \quad (8)$$

The above equations are exact for a *single component plasma* component in equilibrium. In order to derive formulae for the higher moments, additional assumptions and/or approximations are generally necessary. Initially, it is assumed that:

1. the plasma is non-degenerate, ie satisfies *Boltzmann statistics*, so that $\eta \equiv \mu/T \ll -1$.

2. the plasma is weakly coupled so as to allow use of the *Random Phase Approximation* (RPA) for the imaginary part of the dielectric function. This yields, in the case of a non-degenerate plasma [1],

$$\operatorname{Im} \epsilon(\mathbf{q},\omega) \simeq \frac{\sqrt{\pi}}{2} \frac{\Omega_0^2}{T^2} \frac{1}{v^{3/2}} \sinh(\tfrac{1}{2}u) e^{-v/4} e^{-u^2/4v}$$

$$= \sqrt{\frac{2\pi m^3}{T}} \frac{\Omega_0^2}{q^3} \sinh\left(\frac{\omega}{2T}\right) e^{-q^2/8mT} e^{-m\omega^2/2q^2 T}$$

(9)

where $u = \omega/T$ and $v = q^2/2mT$.

3. the *Generalized Plasmon Pole Approximation* relation [1],

$$\mathrm{Im}\left(\frac{-1}{\epsilon(\mathbf{q},\omega)}\right) = F(\mathbf{q})\mathrm{Im}\,\epsilon(\mathbf{q},\omega) + \frac{\pi}{2\omega}\Omega_0^2\left(1 - F(\mathbf{q})\right)\left(\delta(\omega - \Omega_\mathbf{q}) + \delta(\omega + \Omega_\mathbf{q})\right) \quad (10)$$

holds for real frequencies $\Omega_\mathbf{q}$ defined by

$$\epsilon(\mathbf{q}, \Omega_\mathbf{q}) = 0 \quad (11)$$

which represents an undamped plasma *collective mode* with wavenumber $\mathbf{q}$;

and where

$$F(\mathbf{q}) = \frac{1}{\epsilon(\mathbf{q},0)}\left(1 - \frac{(\epsilon(\mathbf{q},0) - 1)\Omega_0^2}{(\epsilon(\mathbf{q},0) - 1)\Omega_\mathbf{q}^2 - \Omega_0^2}\right) \quad (12)$$

## 2 CALCULATION OF MOMENTS OF THE DYNAMIC STRUCTURE FACTOR

### 2.1 Preliminary

The even and odd moments of $S(\mathbf{q},\omega)$ are considered separately. For the even moments, the KMS relation (2) and the fluctuation dissipation theorem (3) yield,

$$\begin{aligned}\mathcal{S}_{2j}(\mathbf{q}) &= \int_{-\infty}^{+\infty}\omega^{2j} S(\mathbf{q},\omega)\,\mathrm{d}\omega \\ &= \int_0^{+\infty}\omega^{2j}\left(S(\mathbf{q},\omega) + S(\mathbf{q},-\omega)\right)\mathrm{d}\omega \\ &= \int_0^{+\infty}\omega^{2j}\left(1 + e^{-\omega/T}\right)S(\mathbf{q},\omega)\,\mathrm{d}\omega \\ &= \frac{q^2}{\pi m \Omega_0^2}\int_0^{+\infty}\omega^{2j}\coth\left(\frac{\omega}{2T}\right)\mathrm{Im}\left(\frac{-1}{\epsilon(\mathbf{q},\omega)}\right)\mathrm{d}\omega\end{aligned} \quad (13)$$

In a similar fashion,

$$S_{2j-1}(\mathbf{q}) = \int_{-\infty}^{+\infty} \omega^{2j-1} S(\mathbf{q},\omega) \, d\omega$$

$$= \int_{0}^{+\infty} \omega^{2j-1} \left( S(\mathbf{q},\omega) - S(\mathbf{q},-\omega) \right) d\omega$$

$$= \int_{0}^{+\infty} \omega^{2j-1} \left( 1 - e^{-\omega/T} \right) S(\mathbf{q},\omega) \, d\omega \qquad (14)$$

$$= \frac{q^2}{\pi m \Omega_0^2} \int_{0}^{+\infty} \omega^{2j-1} \operatorname{Im}\left( \frac{-1}{\epsilon(\mathbf{q},\omega)} \right) d\omega$$

The integrals are thus transformed into ones involving the imaginary part of the inverse dielectric function. Since $\operatorname{Im}\epsilon(\mathbf{q},-\omega) = -\operatorname{Im}\epsilon(\mathbf{q},\omega)$, $\omega \in \mathbb{R}$, these formulae can also be expressed as follows,

$$S_{2j}(\mathbf{q}) = \frac{q^2}{2\pi m \Omega_0^2} \int_{-\infty}^{+\infty} \omega^{2j} \coth\left(\frac{\omega}{2T}\right) \operatorname{Im}\left( \frac{-1}{\epsilon(\mathbf{q},\omega)} \right) d\omega \qquad (15)$$

$$S_{2j-1}(\mathbf{q}) = \frac{q^2}{2\pi m \Omega_0^2} \int_{-\infty}^{+\infty} \omega^{2j-1} \operatorname{Im}\left( \frac{-1}{\epsilon(\mathbf{q},\omega)} \right) d\omega \qquad (16)$$

However, it is more convenient to consider the ongoing calculations in terms of integrations confined to the positive real frequency axis.

## 2.2 Integrals Involving the Dielectric Function

### 2.2.1 Integrals involving Im $\epsilon$

Evaluation of (13) - (14) involves making use of the GPPA formula (10), from which the following integrals emerge

$$\mathcal{R}_{2j}(\mathbf{q}) = \frac{q^2}{\pi m \Omega_0^2} \int_{0}^{+\infty} \omega^{2j} \coth\left(\frac{\omega}{2T}\right) \operatorname{Im}(\epsilon(\mathbf{q},\omega)) \, d\omega \qquad (17)$$

$$\mathcal{R}_{2j-1}(\mathbf{q}) = \frac{q^2}{\pi m \Omega_0^2} \int_{0}^{+\infty} \omega^{2j-1} \operatorname{Im}(\epsilon(\mathbf{q},\omega)) \, d\omega \qquad (18)$$

Using (9), these are given, in the RPA, by

$$\mathfrak{R}_{2j}(\mathbf{q}) = \frac{T^{2j}}{\sqrt{\pi v}} e^{-v/4} \int_0^{+\infty} u^{2j} \cosh\left(\tfrac{1}{2}u\right) e^{-u^2/4v} du$$

$$= \frac{T^{2j}}{\sqrt{\pi v}} e^{-v/4} \int_0^{+\infty} u^{2j} \cosh\left(\tfrac{1}{2}u\right) e^{-u^2/4v} du \qquad (19)$$

$$= \frac{2}{\sqrt{\pi}} \left(4T^2 v\right)^j e^{-v/4} \int_0^{+\infty} x^{2j} \cosh\left(x\sqrt{v}\right) e^{-x^2} dx$$

$$= T^{2j} \frac{\Gamma(2j+1)}{\Gamma(j+1)} v^j f_j(v)$$

$$\mathfrak{R}_{2j-1}(\mathbf{q}) = \frac{T^{2j-1}}{\sqrt{\pi v}} e^{-v/4} \int_0^{+\infty} u^{2j-1} \sinh\left(\tfrac{1}{2}u\right) e^{-u^2/4v} du$$

$$= \frac{T^{2j-1}}{\sqrt{\pi v}} e^{-v/4} \int_0^{+\infty} u^{2j-1} \sinh\left(\tfrac{1}{2}u\right) e^{-u^2/4v} du \qquad (20)$$

$$= \frac{2}{\sqrt{\pi}} \left(2T\sqrt{v}\right)^{2j-1} e^{-v/4} \int_0^{+\infty} x^{2j-1} \sinh\left(x\sqrt{v}\right) e^{-x^2} dx$$

$$= T^{2j-1} \frac{\Gamma(2j+1)}{2\Gamma(j+1)} v^j g_j(v)$$

where

$$f_j(v) = \frac{2^{2j+1}}{\sqrt{\pi}} \frac{\Gamma(j+1)}{\Gamma(2j+1)} e^{-v/4} \int_0^{+\infty} x^{2j} \cosh\left(x\sqrt{v}\right) e^{-x^2} dx$$

$$= e^{-v/4} {}_1F_1\left(j+\tfrac{1}{2};\tfrac{1}{2};\tfrac{1}{4}v\right) \qquad (21)$$

$$\equiv {}_1F_1\left(-j;\tfrac{1}{2};-\tfrac{1}{4}v\right)$$

in which ${}_1F_1(a;b;z)$ denotes the Kummer confluent hypergeometric function [7], and

$$g_j(v) = \frac{2^{2j+1}}{\sqrt{\pi v}} \frac{\Gamma(j+1)}{\Gamma(2j+1)} e^{-v/4} \int_0^{+\infty} x^{2j-1} \sinh(x\sqrt{v}) e^{-x^2} dx$$

$$= e^{-v/4} \,_1F_1\left(j+\tfrac{1}{2}; \tfrac{3}{2}; \tfrac{1}{4}v\right) \tag{22}$$

$$\equiv \,_1F_1\left(1-j; \tfrac{3}{2}; -\tfrac{1}{4}v\right)$$

For integer values of $j \geq 0$, the functions $f_j(x)$ and $g_{j+1}(x)$ are polynomials in $x$ of degree $j$. These polynomials are related to the Hermite polynomials, $H_n$ (see APPENDIX A). Some particular properties of these functions are as follows;

$$f_0(x) = 1$$

$$g_1(x) = 1$$

$$f_j(0) = 1 \tag{23}$$

$$g_j(0) = 1$$

while the function $g_0(x)$ is a special case, and is given by

$$g_0(x) = \frac{2}{\sqrt{x}} \Phi\left(\tfrac{1}{2}\sqrt{x}\right) \tag{24}$$

where $\Phi(x)$ is Dawson's integral,

$$\Phi(x) = e^{-x^2} \int_0^x e^{t^2} dt \tag{25}$$

For small $x$,

$$f_j(x) \sim 1 + \tfrac{1}{2} jx + \mathcal{O}(x^2)$$

$$g_j(x) \sim 1 + \tfrac{1}{6}(j-1)x + \mathcal{O}(x^2) \tag{26}$$

Explicit formulae for $f_j(x)$ and $g_j(x)$ for $j \leq 10$ are given in APPENDIX A.

Using these formulae, the frequency-moments involving the dielectric function are given by:

$$\int_0^{+\infty} \omega^{2j} \coth\left(\frac{\omega}{2T}\right) \operatorname{Im}(\epsilon(\mathbf{q},\omega)) \, d\omega = \frac{\pi m \Omega_0^2}{q^2} \mathcal{R}_{2j}(\mathbf{q}) = \frac{\pi m \Omega_0^2}{q^2} \frac{\Gamma(2j+1)}{\Gamma(j+1)} \left(\frac{q^2 T}{2m}\right)^j f_j\left(\frac{q^2}{2mT}\right)$$
(27)

$$\int_0^{+\infty} \omega^{2j-1} \operatorname{Im}(\epsilon(\mathbf{q},\omega)) \, d\omega = \frac{\pi m \Omega_0^2}{q^2} \mathcal{R}_{2j-1}(\mathbf{q}) = \frac{\pi m \Omega_0^2}{2q^2 T} \frac{\Gamma(2j+1)}{\Gamma(j+1)} \left(\frac{q^2 T}{2m}\right)^j g_j\left(\frac{q^2}{2mT}\right)$$
(28)

which yield, for the first few moments,

$$\int_0^{+\infty} \omega^{-1} \operatorname{Im}(\epsilon(\mathbf{q},\omega)) \, d\omega = \frac{1}{q^2 D^2} \frac{\pi}{\sqrt{v}} \Phi\left(\tfrac{1}{2}\sqrt{v}\right) \tag{29}$$

$$\int_0^{+\infty} \coth\left(\frac{\omega}{2T}\right) \operatorname{Im}(\epsilon(\mathbf{q},\omega)) \, d\omega = \frac{\pi m \Omega_0^2}{q^2} = \frac{\pi T}{q^2 D^2} \tag{30}$$

$$\int_0^{+\infty} \omega \operatorname{Im}(\epsilon(\mathbf{q},\omega)) \, d\omega = \frac{\pi}{2} \Omega_0^2 \tag{31}$$

$$\int_0^{+\infty} \omega^2 \coth\left(\frac{\omega}{2T}\right) \operatorname{Im}(\epsilon(\mathbf{q},\omega)) \, d\omega = \pi \Omega_0^2 T \left(1 + \frac{q^2}{4mT}\right) \tag{32}$$

$$\int_0^{+\infty} \omega^3 \operatorname{Im}(\epsilon(\mathbf{q},\omega)) \, d\omega = \frac{3\pi}{2} \Omega_0^4 q^2 D^2 \left(1 + \frac{q^2}{12mT}\right) \tag{33}$$

in which $D$ is the classical Debye length defined by

$$D^2 = \frac{T}{m \Omega_0^2} \tag{34}$$

and $v = q^2/2mT$. Equation (31) is the well-known conductivity sum rule. Equation (29) is a form of the general compressibility sum rule [1], according to which

$$\int_0^{+\infty} \omega^{-1} \operatorname{Im}(\epsilon(\mathbf{q},\omega)) \, d\omega = \frac{\pi}{2}(\epsilon(\mathbf{q},0)-1) \tag{35}$$

This yields

$$\epsilon(\mathbf{q},0) = 1 + \frac{g_0(v)}{q^2 D^2}$$

$$= 1 + \frac{1}{q^2 D^2} \frac{2}{\sqrt{v}} \Phi\left(\tfrac{1}{2}\sqrt{v}\right) \tag{36}$$

which is a formula obtained, for non-degenerate plasmas, in ref. [1].

The corresponding integrals involving $\operatorname{Im}(-1/\epsilon(\mathbf{q},\omega))$ are given in the next section.

## 2.3  Structure Factor Moments

The moments $\mathcal{S}_n(\mathbf{q})$ of the structure factor can now be deduced with the aid of equations (10), (13), (14), (17), (18), (19) and (20). Hence, for $j \geq 0$,

$$\mathcal{S}_{2j}(\mathbf{q}) = \frac{q^2}{\pi m \Omega_0^2} \int_0^{+\infty} \omega^{2j} \coth\left(\frac{\omega}{2T}\right) \operatorname{Im}\left(\frac{-1}{\epsilon(\mathbf{q},\omega)}\right) d\omega$$

$$= F(\mathbf{q}) \mathcal{R}_{2j}(\mathbf{q}) + (1-F(\mathbf{q})) \frac{q^2}{2m} \Omega_\mathbf{q}^{2j-1} \coth\left(\frac{\Omega_\mathbf{q}}{2T}\right)$$

$$= F(\mathbf{q}) \frac{\Gamma(2j+1)}{\Gamma(j+1)} \left(\frac{q^2 T}{2m}\right)^j f_j\left(\frac{q^2}{2mT}\right) + (1-F(\mathbf{q})) \frac{q^2}{2m} \Omega_\mathbf{q}^{2j-1} \coth\left(\frac{\Omega_\mathbf{q}}{2T}\right)$$

$$= T^{2j} \left(\frac{q^2}{mT}\right)^j \left( F(\mathbf{q}) \frac{\Gamma(2j+1)}{2^j \Gamma(j+1)} f_j\left(\frac{q^2}{2mT}\right) + (1-F(\mathbf{q})) \left(\frac{1}{q^2 D^2} \frac{\Omega_\mathbf{q}^2}{\Omega_0^2}\right)^{j-1} \left(\frac{\Omega_\mathbf{q}}{2T}\right) \coth\left(\frac{\Omega_\mathbf{q}}{2T}\right) \right)$$

(37)

for the even moments; and

$$\mathcal{S}_{2j-1}(\mathbf{q}) = \frac{q^2}{\pi m \Omega_0^2} \int_0^{+\infty} \omega^{2j-1} \operatorname{Im}\left(\frac{-1}{\epsilon(\mathbf{q},\omega)}\right) d\omega$$

$$= F(\mathbf{q}) \mathcal{R}_{2j-1}(\mathbf{q}) + (1 - F(\mathbf{q})) \frac{q^2}{2m} \Omega_{\mathbf{q}}^{2j-2}$$

$$= F(\mathbf{q}) \frac{\Gamma(2j+1)}{\Gamma(j+1)} \frac{1}{2T} \left(\frac{q^2 T}{2m}\right)^j g_j\left(\frac{q^2}{2mT}\right) + (1 - F(\mathbf{q})) \frac{q^2}{2m} \Omega_{\mathbf{q}}^{2j-2}$$

$$= \frac{T^{2j-1}}{2} \left(\frac{q^2}{mT}\right)^j \left( F(\mathbf{q}) \frac{\Gamma(2j+1)}{2^j \Gamma(j+1)} g_j\left(\frac{q^2}{2mT}\right) + (1 - F(\mathbf{q})) \left(\frac{1}{q^2 D^2} \frac{\Omega_{\mathbf{q}}^2}{\Omega_0^2}\right)^{j-1} \right)$$

(38)

for the odd moments. The relationship between successive moments can be expressed as

$$\mathcal{S}_{2j} - 2T\mathcal{S}_{2j-1} = T^{2j} \left(\frac{q^2}{mT}\right)^j \left( \begin{array}{c} F(\mathbf{q}) \frac{\Gamma(2j+1)}{2^j \Gamma(j+1)} \left( f_j\left(\frac{q^2}{2mT}\right) - g_j\left(\frac{q^2}{2mT}\right) \right) \\ + (1 - F(\mathbf{q})) \left(\frac{1}{q^2 D^2} \frac{\Omega_{\mathbf{q}}^2}{\Omega_0^2}\right)^{j-1} \left( \frac{\Omega_{\mathbf{q}}}{2T} \coth\left(\frac{\Omega_{\mathbf{q}}}{2T}\right) - 1 \right) \end{array} \right)$$

(39)

Using these formulae, the first few moments, corresponding to (29) - (33), are found to be given as follows, with use being made, where appropriate, of (12) and (36),

$$\mathcal{S}_{-1}(\mathbf{q}) = F(\mathbf{q})\frac{g_0(v)}{2T} + (1-F(\mathbf{q}))\frac{q^2}{2m\Omega_\mathbf{q}^2}$$

$$= \frac{q^2 D^2}{2T}\left( F(\mathbf{q})(\epsilon(\mathbf{q},0)-1) + (1-F(\mathbf{q}))\frac{\Omega_0^2}{\Omega_\mathbf{q}^2} \right)$$

$$= \frac{q^2 D^2}{2T}\left( 1 - \frac{1}{\epsilon(\mathbf{q},0)} \right)$$

$$= \frac{1}{2T}\frac{g_0(v)q^2 D^2}{g_0(v)+q^2 D^2}$$

$$= \frac{1}{2T}\frac{g_0(v)}{\epsilon(\mathbf{q},0)} \tag{40}$$

$$\mathcal{S}_0(\mathbf{q}) = F(\mathbf{q}) + (1-F(\mathbf{q}))\frac{q^2}{2m\Omega_\mathbf{q}}\coth\left(\frac{\Omega_\mathbf{q}}{2T}\right)$$

$$= F(\mathbf{q}) + (1-F(\mathbf{q}))\frac{q^2 D^2}{g_0(v)}\frac{\Omega_0^2}{\Omega_\mathbf{q}^2} + (1-F(\mathbf{q}))\frac{q^2 D^2}{g_0(v)}\frac{\Omega_0^2}{\Omega_\mathbf{q}^2}\left( g_0(v)\frac{\Omega_\mathbf{q}}{2T}\coth\left(\frac{\Omega_\mathbf{q}}{2T}\right) - 1 \right)$$

$$= \frac{q^2 D^2}{g_0(v)+q^2 D^2}\left( 1 + \frac{g_0(v)\Omega_0^2}{g_0(v)\Omega_\mathbf{q}^2 - \Omega_0^2 q^2 D^2}\left( g_0(v)\frac{\Omega_\mathbf{q}}{2T}\coth\left(\frac{\Omega_\mathbf{q}}{2T}\right) - 1 \right) \right)$$

$$= \frac{1}{\epsilon(\mathbf{q},0)}\left( 1 + \frac{(\epsilon(\mathbf{q},0)-1)\Omega_0^2}{(\epsilon(\mathbf{q},0)-1)\Omega_\mathbf{q}^2 - \Omega_0^2}\left( g_0(v)\frac{\Omega_\mathbf{q}}{2T}\coth\left(\frac{\Omega_\mathbf{q}}{2T}\right) - 1 \right) \right)$$

$$\tag{41}$$

$$\mathcal{S}_1(\mathbf{q}) = \frac{q^2}{2m} \tag{42}$$

$$\mathcal{S}_2(\mathbf{q}) = \frac{q^2 T}{m}\left(1 + F(\mathbf{q})\frac{q^2}{4mT} + (1-F(\mathbf{q}))\left(\frac{\Omega_\mathbf{q}}{2T}\coth\left(\frac{\Omega_\mathbf{q}}{2T}\right) - 1\right)\right) \qquad (43)$$

$$\mathcal{S}_3(\mathbf{q}) = 2T^3\left(\frac{q^2}{2mT}\right)^2\left(F(\mathbf{q})\left(3 + \frac{q^2}{4mT}\right) + (1-F(\mathbf{q}))\frac{1}{q^2 D^2}\frac{\Omega_\mathbf{q}^2}{\Omega_0^2}\right) \qquad (44)$$

Equation (40) expresses the screening sum rule for a single-component plasma [1]. Equation (41) yields the GPPA static structure factor [1] in accordance with the elastic sum rule (6), for a single-component non-degenerate plasma component. Equation (42) expresses the f-sum rule, which is an exact result for non-relativistic systems.

### 2.3.1 Semiclassical Approximations

Hot plasmas, those for which $T \gg \Omega_0$, can generally be treated classically or semiclassically. Quantities such as $\Omega_\mathbf{q}/T$ are first order *semiclassical* quantities (being first order in $\hbar$) while $q^2/2mT$ is a second-order semiclassical quantity. The lowest-order semiclassical approximations for the structure factor moments are therefore

$$\mathcal{S}_{-1}(\mathbf{q}) \simeq \frac{1}{2T}\left(\frac{q^2 D^2}{1 + q^2 D^2}\right) \qquad (45)$$

$$\mathcal{S}_0(\mathbf{q}) \simeq \frac{q^2 D^2}{1 + q^2 D^2} \qquad (46)$$

$$\mathcal{S}_1(\mathbf{q}) = \frac{q^2}{2m} \qquad (47)$$

$$\mathcal{S}_2(\mathbf{q}) \simeq \frac{q^2 T}{m} \qquad (48)$$

$$\mathcal{G}_3(\mathbf{q}) \simeq 2T^3 \left(\frac{q^2}{2mT}\right)^2 \left(3F(\mathbf{q}) + (1-F(\mathbf{q}))\frac{\Omega_q^2}{\Omega_0^2}\frac{1}{q^2 D^2}\right) \tag{49}$$

$$= \frac{\Omega_0^4}{2T} q^2 D^2 \left(3F(\mathbf{q})q^2 D^2 + (1-F(\mathbf{q}))\frac{\Omega_q^2}{\Omega_0^2}\right)$$

Equation (45) is the *Debye-Hückel* screening sum rule, whereby, with the aid of (12) and (35),

$$F(\mathbf{q}) \simeq 1 - \frac{1}{1+q^2 D^2}\left(\frac{\Omega_q^2}{\Omega_q^2 - \Omega_0^2 q^2 D^2}\right) \tag{50}$$

Equation (46) is the elastic sum rule for the Debye-Hückel static structure factor. Equation (47) is the f-sum rule, which retains its unapproximated form.

In general, it follows from (39) that the successive moments, $\mathcal{G}_{2j-1}(\mathbf{q})$ and $\mathcal{G}_{2j}(\mathbf{q})$, are always both $\mathcal{O}(\hbar^{2j})$ and are related by

$$\mathcal{G}_{2j}(\mathbf{q}) \simeq 2T\mathcal{G}_{2j-1}(\mathbf{q}) \tag{51}$$

so, for example, from (49),

$$\mathcal{G}_4(\mathbf{q}) \simeq \Omega_0^4 q^2 D^2 \left(3F(\mathbf{q})q^2 D^2 + (1-F(\mathbf{q}))\frac{\Omega_q^2}{\Omega_0^2}\right) \tag{52}$$

## 3　DEGENERATE PLASMAS

### 3.1　Integrals Involving the Dielectric Function

The generalisation of the above to degenerate and partially degenerate plasmas ($\eta \not\ll 0$) involves using the more general formula [1],

$$\operatorname{Im}\epsilon(\mathbf{q},\omega) \simeq \frac{\pi}{4}\frac{\Omega_0^2}{T^2}\frac{1}{v^{3/2}}\frac{1}{I_{1/2}(\eta)}\frac{\sinh\left(\tfrac{1}{2}u\right)}{1+\exp\left(u^2/4v + v/4 - \eta\right)} \tag{53}$$

in place of (9), where

$$I_j(x) = \int_0^\infty \frac{y^j}{1+\exp(y-x)}\,dy \tag{54}$$

denotes a Fermi integral. The formulae (17) - (18) then become

$$\mathcal{R}_{2j}(\mathbf{q}) = \frac{q^2 T^{2j-1}}{4m} \frac{1}{v^{3/2}} \frac{1}{I_{1/2}(\eta)} \int_0^{+\infty} \frac{\cosh(\tfrac{1}{2}u)}{1+\exp(u^2/4v+v/4-\eta)} u^{2j} \, du$$

$$= \frac{1}{2}(2T)^{2j} v^j \frac{1}{I_{1/2}(\eta)} \int_0^{+\infty} \frac{\cosh(\sqrt{vy})}{1+\exp(y+v/4-\eta)} y^{j-1/2} \, dy \tag{55}$$

$$\mathcal{R}_{2j-1}(\mathbf{q}) = \frac{q^2 T^{2j-2}}{4m} \frac{1}{v^{3/2}} \frac{1}{I_{1/2}(\eta)} \int_0^{+\infty} \frac{\sinh(\tfrac{1}{2}u)}{1+\exp(u^2/4v+v/4-\eta)} u^{2j-1} \, du$$

$$= \frac{1}{2}(2T)^{2j-1} v^{j-1/2} \frac{1}{I_{1/2}(\eta)} \int_0^{+\infty} \frac{\sinh(\sqrt{vy})}{1+\exp(y+v/4-\eta)} y^{j-1} \, dy \tag{56}$$

These integrals may be evaluated by expanding the hyperbolic functions as power series of their arguments, while making use of (54). This procedure yields the moments of the imaginary part of the dielectric function as infinite series of Fermi functions, as follows

$$\mathcal{R}_{2j}(\mathbf{q}) = \frac{1}{2}(2T)^{2j} \frac{1}{I_{1/2}(\eta)} \sum_{k=0}^{\infty} \frac{v^{k+j}}{\Gamma(2k+1)} I_{k+j-1/2}(\eta-v/4) \tag{57}$$

$$\mathcal{R}_{2j-1}(\mathbf{q}) = \frac{1}{2}(2T)^{2j-1} \frac{1}{I_{1/2}(\eta)} \sum_{k=0}^{\infty} \frac{v^{k+j}}{\Gamma(2k+2)} I_{k+j-1/2}(\eta-v/4) \tag{58}$$

These series are absolutely convergent, and provide a basis for numerical approximations.

### 3.2 Long-Wavelength Approximations for Degenerate Plasmas

If it can be assumed that $q^2 \ll 2mT$ then $v \ll 1$, which leads to the semiclassical forms of (57) - (58) as follows

$$\mathcal{R}_{2j}(\mathbf{q}) \simeq \frac{1}{2}(2T)^{2j} v^j \frac{I_{j-1/2}(\eta)}{I_{1/2}(\eta)} \tag{59}$$

$$\mathcal{R}_{2j-1}(\mathbf{q}) \simeq \frac{1}{2}(2T)^{2j-1} v^j \frac{I_{j-1/2}(\eta)}{I_{1/2}(\eta)} \tag{60}$$

which satisfy

$$\mathcal{R}_{2j}(\mathbf{q}) \simeq 2T \mathcal{R}_{2j-1}(\mathbf{q}) \tag{61}$$

For the first few moments, this yields

$$\mathcal{R}_{-1}(\mathbf{q}) \simeq \frac{1}{4T} \frac{I_{-1/2}(\eta)}{I_{1/2}(\eta)} \equiv \frac{R_0(\eta)}{2T} \tag{62}$$

$$\mathcal{R}_0(\mathbf{q}) \simeq \frac{1}{2} \frac{I_{-1/2}(\eta)}{I_{1/2}(\eta)} \equiv \frac{I'_{1/2}(\eta)}{I_{1/2}(\eta)} \equiv R_0(\eta) \tag{63}$$

$$\mathcal{R}_1(\mathbf{q}) = \frac{q^2}{2m} \tag{64}$$

$$\mathcal{R}_2(\mathbf{q}) \simeq \frac{q^2 T}{m} \tag{65}$$

$$\mathcal{R}_3(\mathbf{q}) \simeq \frac{1}{2}(2T)^3 v^2 \frac{I_{3/2}(\eta)}{I_{1/2}(\eta)} = \left(\frac{q^2}{m}\right)^2 \langle \varepsilon \rangle \tag{66}$$

$$\mathcal{R}_4(\mathbf{q}) \simeq \frac{1}{2}(2T)^4 v^2 \frac{I_{3/2}(\eta)}{I_{1/2}(\eta)} = 2T \left(\frac{q^2}{m}\right)^2 \langle \varepsilon \rangle \tag{67}$$

where $\langle \varepsilon \rangle = T I_{3/2}(\eta)/I_{1/2}(\eta)$ is the expectation one-body energy [8]. Note however, that in the high-density (degenerate) regime, it is generally not appropriate to assume that $T \gg \Omega_{\mathbf{q}}$. Therefore, the prevailing assumption that $q^2 \ll 2mT$ amounts to a restriction on $q$ rather than defining a consistent semiclassical approximation associated with a particular physical regime of the material. Accordingly, in the following, it is neither appropriate nor necessary to assume that the plasma frequency is small compared with the temperature.

Equation (62) modifies the screening sum rule so that, for small $q$, we now have [1]

$$\epsilon(\mathbf{q},0) \simeq 1 + \frac{R_0}{q^2 D^2} = 1 + \frac{1}{q^2 \tilde{D}^2} \tag{68}$$

where $\tilde{D} = D/\sqrt{R_0}$ is the degeneracy-modified Debye length. To lowest order in $q$, the 'zeroth' moment of the dielectric function is then given (30) but with $D$ replaced by $\tilde{D}$; the moments (31) and (32) are unaffected by degeneracy (In particular, the f-sum rule, (31), remains exact.) while the following moments, (33) and (34), are modified by a factor of $2\langle\varepsilon\rangle/3T$, which is unity in the non-degenerate limit.

The moments of the structure factor can now be calculated in accordance with, eg (37) - (38),

$$S_{2j}(\mathbf{q}) = F(\mathbf{q})\mathcal{H}_{2j}(\mathbf{q}) + (1-F(\mathbf{q}))\frac{q^2}{2m}\Omega_{\mathbf{q}}^{2j-1}\coth\left(\frac{\Omega_{\mathbf{q}}}{2T}\right)$$

$$= F(\mathbf{q})\mathcal{H}_{2j}(\mathbf{q}) + (1-F(\mathbf{q}))q^2 D^2 \Omega_0^{2j}\left(\frac{\Omega_{\mathbf{q}}^2}{\Omega_0^2}\right)^{j-1}\left(\frac{\Omega_{\mathbf{q}}}{2T}\coth\left(\frac{\Omega_{\mathbf{q}}}{2T}\right)\right) \quad (69)$$

$$S_{2j-1}(\mathbf{q}) = F(\mathbf{q})\mathcal{H}_{2j-1}(\mathbf{q}) + \frac{1}{2T}(1-F(\mathbf{q}))q^2 D^2 \Omega_0^{2j}\left(\frac{\Omega_{\mathbf{q}}^2}{\Omega_0^2}\right)^{j-1} \quad (70)$$

which yield for the first few,

$$S_{-1}(\mathbf{q}) \simeq \frac{R_0(\eta)}{2T}\left(F(\mathbf{q}) + (1-F(\mathbf{q}))q^2\tilde{D}^2\left(\frac{\Omega_0^2}{\Omega_{\mathbf{q}}^2}\right)\right)$$

$$= \frac{1}{2T_B}\frac{q^2\tilde{D}^2}{1+q^2\tilde{D}^2} \quad (71)$$

$$= \frac{1}{2T_B}\frac{1}{\epsilon(\mathbf{q},0)}$$

$$S_0(\mathbf{q}) \simeq R_0(\eta)\left(\frac{q^2\tilde{D}^2}{1+q^2\tilde{D}^2} + (1-F(\mathbf{q}))q^2\tilde{D}^2\left(\frac{\Omega_0^2}{\Omega_{\mathbf{q}}^2}\right)\left(\frac{\Omega_{\mathbf{q}}}{2T}\coth\left(\frac{\Omega_{\mathbf{q}}}{2T}\right)-1\right)\right)$$

$$= R_0(\eta)\frac{q^2\tilde{D}^2}{1+q^2\tilde{D}^2}\left(1+\left(\frac{\Omega_0^2}{\Omega_{\mathbf{q}}^2-\Omega_0^2 q^2\tilde{D}^2}\right)\left(\frac{\Omega_{\mathbf{q}}}{2T}\coth\left(\frac{\Omega_{\mathbf{q}}}{2T}\right)-1\right)\right) \quad (72)$$

$$= \frac{R_0(\eta)}{\epsilon(\mathbf{q},0)}\left(1+\left(\frac{\Omega_0^2}{\Omega_{\mathbf{q}}^2-\Omega_0^2 q^2\tilde{D}^2}\right)\left(\frac{\Omega_{\mathbf{q}}}{2T}\coth\left(\frac{\Omega_{\mathbf{q}}}{2T}\right)-1\right)\right)$$

$$\mathcal{S}_1(\mathbf{q}) = \frac{q^2}{2m} \tag{73}$$

$$\mathcal{S}_2(\mathbf{q}) \simeq \frac{q^2 T}{m}\left(1 + (1-F(\mathbf{q}))\left(\frac{\Omega_q}{2T}\coth\left(\frac{\Omega_q}{2T}\right) - 1\right)\right)$$

$$= \frac{q^2 T}{m}\left(1 + \left(\frac{1}{1+q^2\tilde{D}^2}\right)\left(\frac{\Omega_q^2}{\Omega_q^2 - \Omega_0^2 q^2 \tilde{D}^2}\right)\left(\frac{\Omega_q}{2T}\coth\left(\frac{\Omega_q}{2T}\right) - 1\right)\right)$$

$$= R_0(\eta)\Omega_0^2\left(\frac{1}{\epsilon(\mathbf{q},0)-1} + \frac{1}{\epsilon(\mathbf{q},0)}\left(\frac{\Omega_q^2}{\Omega_q^2 - \Omega_0^2 q^2 \tilde{D}^2}\right)\left(\frac{\Omega_q}{2T}\coth\left(\frac{\Omega_q}{2T}\right) - 1\right)\right)$$

(74)

$$\mathcal{S}_3(\mathbf{q}) \simeq F(\mathbf{q})\left(\frac{q^2}{m}\right)^2 \langle\varepsilon\rangle + \frac{1}{2T}(1-F(\mathbf{q}))q^2 D^2 \Omega_0^4 \left(\frac{\Omega_q^2}{\Omega_0^2}\right)$$

$$= \frac{\Omega_0^4}{2T}q^2 D^2\left(2\frac{\langle\varepsilon\rangle}{T}F(\mathbf{q})q^2 D^2 + (1-F(\mathbf{q}))\frac{\Omega_q^2}{\Omega_0^2}\right) \tag{75}$$

$$= \frac{\Omega_0^4}{2T_B}q^2\tilde{D}^2\left(2\frac{\langle\varepsilon\rangle}{T_B}F(\mathbf{q})q^2\tilde{D}^2 + (1-F(\mathbf{q}))\frac{\Omega_q^2}{\Omega_0^2}\right)$$

$$\mathcal{S}_4(\mathbf{q}) \simeq 2TF(\mathbf{q})\left(\frac{q^2}{m}\right)^2\langle\varepsilon\rangle + (1-F(\mathbf{q}))q^2 D^2 \Omega_0^4\left(\frac{\Omega_q^2}{\Omega_0^2}\right)\left(\frac{\Omega_q}{2T}\coth\left(\frac{\Omega_q}{2T}\right)\right)$$

(76)

$$= R_0(\eta)\Omega_0^4 q^2 \tilde{D}^2\left(2\frac{\langle\varepsilon\rangle}{T_B}F(\mathbf{q})q^2\tilde{D}^2 + (1-F(\mathbf{q}))\frac{\Omega_q^2}{\Omega_0^2}\left(\frac{\Omega_q}{2T}\coth\left(\frac{\Omega_q}{2T}\right)\right)\right)$$

where $T_B = T/R_0$. Equations (71) - (73) respectively are small–$q$ expressions of the screening sum rule, elastic sum rule and f- sum rule. Equation (72) yields the elastic sum rule for a one-component plasma in agreement with formulae given in ref. [1]. Equations (74) - (76) give the higher moments and are consistent with the corresponding formulae given earlier in this article, in the case of non-degenerate plasmas.

In the limit of extreme degeneracy ($\eta \gg 1$): $T_B \sim \tfrac{2}{3}\eta T$, $\langle \varepsilon \rangle \sim \tfrac{3}{5}\eta T$, $R_0 \sim 3/2\eta$.

## 4 CONCLUSIONS

General formulae for the integer moments of the dynamic structure factor of a non-relativistic non-degenerate weakly-coupled single-component quantum plasma have been derived in the GPPA + RPA approximations. Exact sum rules are recovered where appropriate and formulae for other low-order moments take the form of simple expressions. The semiclassical and classical limits are considered and yield the Debye-Hückel formulae for the screening and elastic sum rules and the static structure factor.

These formulae are valid for all wavelength scales for non-degenerate single-component plasmas. The generalisation to degenerate plasmas is obtained in the limit of small wavenumbers, such that $q^2 \ll 2mT$, only. Otherwise the integrals cannot be expressed in terms of simple functions and are instead provided as infinite series of Fermi integrals.

The generalizations to multicomponent plasmas are straightforward, if a little more complicated, and can be derived using the multicomponent forms of the generalised plasmon pole approximation given in ref. [1]. These are probably best considered on a case-by-case basis.

The results are also applicable to some strongly-coupled regimes. For example, setting $F(\mathbf{q})$ equal to zero in equations (37) and (38) or (69) and (70) yields the corresponding results in the simple plasmon pole approximation [1]. Otherwise a suitable model for the imaginary part of the dielectric function, would need be used in place of (9).

## APPENDIX A: MATHEMATICAL FORMULAE

### A.1  Explicit formulae for the functions $f_j(x)$ for $j \leq 10$

Explicit formulae for the functions $f_j(x)$, $j \in \mathbb{N}$ defined by (21), for $0 \leq j \leq 10$ are as follows [9]:

$$f_0(x) = 1 \tag{77}$$

$$f_1(x) = 1 + \tfrac{1}{2}x \tag{78}$$

$$f_2(x) = 1 + x + \tfrac{1}{12}x^2 \tag{79}$$

$$f_3(x) = 1 + \tfrac{3}{2}x + \tfrac{1}{4}x^2 + \tfrac{1}{120}x^3 \tag{80}$$

$$f_4(x) = 1 + 2x + \tfrac{1}{2}x^2 + \tfrac{1}{30}x^3 + \tfrac{1}{1680}x^4 \tag{81}$$

$$f_5(x) = 1 + \tfrac{5}{2}x + \tfrac{5}{6}x^2 + \tfrac{1}{12}x^3 + \tfrac{1}{336}x^4 + \tfrac{1}{30240}x^5 \tag{82}$$

$$f_6(x) = 1 + 3x + \tfrac{5}{4}x^2 + \tfrac{1}{6}x^3 + \tfrac{1}{112}x^4 + \tfrac{1}{5040}x^5 + \tfrac{1}{665280}x^6 \tag{83}$$

$$f_7(x) = 1 + \tfrac{7}{2}x + \tfrac{7}{4}x^2 + \tfrac{7}{24}x^3 + \tfrac{1}{48}x^4 + \tfrac{1}{1440}x^5 + \tfrac{1}{95040}x^6 + \tfrac{1}{17297280}x^7 \tag{84}$$

$$f_8(x) = 1 + 4x + \tfrac{7}{3}x^2 + \tfrac{7}{15}x^3 + \tfrac{1}{24}x^4 + \tfrac{1}{540}x^5 + \tfrac{1}{23760}x^6 + \tfrac{1}{2162160}x^7 + \tfrac{1}{518918400}x^8 \tag{85}$$

$$f_9(x) = 1 + \tfrac{9}{2}x + 3x^2 + \tfrac{7}{10}x^3 + \tfrac{3}{40}x^4 + \tfrac{1}{240}x^5 + \tfrac{1}{7920}x^6 + \tfrac{1}{480480}x^7 + \tfrac{1}{57657600}x^8 + \tfrac{1}{17643225600}x^9 \tag{86}$$

$$f_{10}(x) = 1 + 5x + \tfrac{15}{4}x^2 + x^3 + \tfrac{1}{8}x^4 + \tfrac{1}{120}x^5 + \tfrac{1}{3168}x^6 + \tfrac{1}{144144}x^7 + \tfrac{1}{11531520}x^8 + \tfrac{1}{1764322560}x^9 + \tfrac{1}{670442572800}x^{10} \tag{87}$$

In terms of the Hermite polynomials $H_n$,

$$f_j(x) = \frac{\Gamma(j+1)}{\Gamma(2j+1)}(-1)^j H_{2j}\left(\tfrac{1}{2}i\sqrt{x}\right) \tag{88}$$

## A.2 Explicit formulae for the functions $g_j(x)$ for $j \leq 10$

Explicit formulae for the functions, $g_j(x)$, $j \in \mathbb{N}$, defined by (22), for $0 \leq j \leq 10$, are as follows [9]:

$$g_0(x) = \frac{2}{\sqrt{x}} \Phi\left(\frac{\sqrt{x}}{2}\right) \tag{89}$$

$$g_1(x) = 1 \tag{90}$$

$$g_2(x) = 1 + \tfrac{1}{6} x \tag{91}$$

$$g_3(x) = 1 + \tfrac{1}{3} x + \tfrac{1}{60} x^2 \tag{92}$$

$$g_4(x) = 1 + \tfrac{1}{2} x + \tfrac{1}{20} x^2 + \tfrac{1}{840} x^3 \tag{93}$$

$$g_5(x) = 1 + \tfrac{2}{3} x + \tfrac{1}{10} x^2 + \tfrac{1}{210} x^3 + \tfrac{1}{15120} x^4 \tag{94}$$

$$g_6(x) = 1 + \tfrac{5}{6} x + \tfrac{1}{6} x^2 + \tfrac{1}{84} x^3 + \tfrac{1}{3024} x^4 + \tfrac{1}{332640} x^5 \tag{95}$$

$$g_7(x) = 1 + x + \tfrac{1}{4} x^2 + \tfrac{1}{42} x^3 + \tfrac{1}{1008} x^4 + \tfrac{1}{55440} x^5 + \tfrac{1}{8648640} x^6 \tag{96}$$

$$g_8(x) = 1 + \tfrac{7}{6} x + \tfrac{7}{20} x^2 + \tfrac{1}{24} x^3 + \tfrac{1}{432} x^4 + \tfrac{1}{15840} x^5 + \tfrac{1}{1235520} x^6 + \tfrac{1}{259459200} x^7 \tag{97}$$

$$g_9(x) = 1 + \tfrac{4}{3} x + \tfrac{7}{15} x^2 + \tfrac{1}{15} x^3 + \tfrac{1}{216} x^4 + \tfrac{1}{5940} x^5 + \tfrac{1}{308880} x^6 + \tfrac{1}{32432400} x^7 + \tfrac{1}{8821612800} x^8 \tag{98}$$

$$g_{10}(x) = 1 + \tfrac{3}{2} x + \tfrac{3}{5} x^2 + \tfrac{1}{10} x^3 + \tfrac{1}{120} x^4 + \tfrac{1}{2640} x^5 + \tfrac{1}{102960} x^6 + \tfrac{1}{7207200} x^7 + \tfrac{1}{980179200} x^8 + \tfrac{1}{335221286400} x^9$$

$$\tag{99}$$

In terms of the Hermite polynomials $H_n$,

$$g_j(x) = \frac{\Gamma(j)}{\Gamma(2j)} (-1)^{j+1} \frac{1}{i\sqrt{x}} H_{2j-1}\left(\tfrac{1}{2} i \sqrt{x}\right), \quad j \geq 1 \tag{100}$$

## APPENDIX B: LIST OF SYMBOLS

### B.1 List of Symbols Used for Mathematical and Physical Quantities

$D$     Classical Debye length defined by (34).

$\tilde{D}$     $= D/\sqrt{R_0}$ .

$e$     Electric charge of particle species.

$F(\mathbf{q})$     GPPA coefficients appearing in (10) and given generally by (12).

${}_1F_1(a;b;c)$ Kummer confluent hypergeometric function [7].

$f_j(x)$     (Polynomial) function defined by (21).

$g_j(x)$     Function defined by (22).

$\hat{H}$     Hamiltonian operator.

$H_n(x)$ Hermite polynomial [7].

$\hbar$     Planck's constant $(= h/2\pi)$. (Set equal to unity in formulae as given.)

$I_j(x)$     Fermi integral, defined by (54).

$j$     Integer, expressing order of moment.

$k$     Boltzmann's constant. (Set equal to unity.)

$m$     Particle mass.

$n$     Mean particle density, or general integer.

$\mathbb{N}$     The set of integers.

$\mathbf{q}$     Wavevector variable.

$q$     Wavenumber variable $= |\mathbf{q}|$.

$R_0$     $R_0(\eta) = \dfrac{1}{2}\dfrac{I_{-1/2}(\eta)}{I_{1/2}(\eta)} = \dfrac{I'_{1/2}(\eta)}{I_{1/2}(\eta)}$

$\mathbf{r}$     Coordinate vector.

$\mathcal{R}_n(\mathbf{q})$ Integral, related to $n^{\text{th}}$ moment of $\operatorname{Im}\epsilon(\mathbf{q},\omega)$ defined by (17) - (18).

$\mathbb{R}$     The set of all real numbers.

$S(\mathbf{q},\omega)$ Dynamic structure factor (1).

$S(\mathbf{q})$     Static structure factor $= \mathcal{S}_0(\mathbf{q})$.

$$\mathcal{S}_n(\mathbf{q}) = \int_{-\infty}^{+\infty} \omega^n S(\mathbf{q},\omega) \, d\omega$$

$T$     Temperature.

$T_B$     $= T/R_0$

$t$     Time

$u$     $= \omega/T$

$v$     $= q^2/2mT$

$\mathfrak{V}$     Volume

$x$     General variable.

$\Gamma(z)$     Euler gamma function [7].

$\delta(x)$     Dirac delta function.

$\varepsilon$     Particle energy.

$\epsilon_0$     Permittivity of free space.

$\epsilon(\mathbf{q},\omega)$ Longitudinal dielectric function.

$\eta$     $= \mu/T$ = degeneracy parameter.

$\mu$     Chemical potential.

$\hat{\rho}_\mathbf{r}$     Density operator in coordinate space.

$\hat{\rho}_\mathbf{q}$     Density operator in momentum-space.

$\Phi(x)$     Dawson's integral, as defined by (25).

$\Omega_\mathbf{q}$     Plasma frequency corresponding to wavevector mode $\mathbf{q}$, where $\mathbf{q} = \mathbf{0}$ gives the limiting plasma frequency (4).)

$\omega$     Frequency variable.

This page is intentionally left blank

This page is intentionally left blank